\documentclass[traditabstract]{aa}

\usepackage{natbib} 
\usepackage{color}
\bibpunct{(}{)}{;}{a}{}{,} 
\usepackage{graphicx}
\usepackage{amsmath}
\usepackage{txfonts}
\usepackage{xspace}
\usepackage{amssymb}
\newcommand{\teff}{\mbox{$T_{\rm *, eff}$}\xspace}
\newcommand{\logg}{\mbox{$\log g$}\xspace}
\newcommand{\vsini}{\mbox{$v \sin i_{*}$}\xspace}
\newcommand{\mictrb}{\mbox{$\xi_{\rm t}$}\xspace}
\newcommand{\mactrb}{\mbox{$v_{\rm mac}$}\xspace}

\newcommand{\kms}{\mbox{km\,s$^{-1}$}\xspace}

\newcommand{\mjup}{\mbox{$M_{\rm Jup}$}\xspace}
\newcommand{\rjup}{\mbox{$R_{\rm Jup}$}\xspace}

\newcommand{\rstar}{\mbox{$R_{*}$}\xspace}

\newcommand{\msol}{\mbox{$M_\odot$}\xspace}
\newcommand{\rsol}{\mbox{$R_\odot$}\xspace}

\begin{document}
   \title{WASP-104b and WASP-106b: two transiting hot Jupiters in 1.75-day and 9.3-day orbits}


\authorrunning{A. M. S. Smith \textit{et al.}}
\titlerunning{WASP-104b and WASP-106b}

\author{
	A.~M.~S.~Smith\inst{\ref{inst1},\ref{inst2}}\and
          D.~R.~Anderson\inst{\ref{inst2}}\and
          D.~J.~Armstrong\inst{\ref{inst3}}\and
          S.~C.~C.~Barros\inst{\ref{inst4}}\and
          A.~S.~Bonomo\inst{\ref{inst5}}\and					
          F.~Bouchy\inst{\ref{inst4}}\and
	 D.~J.~A.~Brown\inst{\ref{inst3}}\and
          A.~Collier Cameron\inst{\ref{inst6}}\and
          L.~Delrez\inst{\ref{inst7}}\and
	 F.~Faedi\inst{\ref{inst3}}\and
          M.~Gillon\inst{\ref{inst7}}\and
	 Y.~G\'{o}mez Maqueo Chew\inst{\ref{inst3}}\and
          G.~H\'{e}brard\inst{\ref{inst8},\ref{inst9}}\and
          E.~Jehin\inst{\ref{inst7}}\and
          M.~Lendl\inst{\ref{inst7},\ref{inst10}}\and
          T.~M.~Louden\inst{\ref{inst3}}\and
          P.~F.~L.~Maxted\inst{\ref{inst2}}\and
          G.~Montagnier\inst{\ref{inst8},\ref{inst9}}\and
          M.~Neveu-VanMalle\inst{\ref{inst10},\ref{inst11}}\and
          H.~P.~Osborn\inst{\ref{inst3}}\and
          F.~Pepe\inst{\ref{inst10}}\and
          D.~Pollacco\inst{\ref{inst3}}\and
          D.~Queloz\inst{\ref{inst10},\ref{inst11}}\and
          J.~W.~Rostron\inst{\ref{inst3}}\and
          D.~Segransan\inst{\ref{inst10}}\and
          B.~Smalley\inst{\ref{inst2}}\and
          A.~H.~M.~J.~Triaud\inst{\ref{inst10},\ref{inst12}}\and
          O.~D.~Turner\inst{\ref{inst2}}\and
          S.~Udry\inst{\ref{inst10}}\and
          S.~R.~Walker\inst{\ref{inst3}}\and
          R.~G.~West\inst{\ref{inst3}}\and
          P.~J.~Wheatley\inst{\ref{inst3}}
                    }

\institute{
N. Copernicus Astronomical Centre, Polish Academy of Sciences, Bartycka 18, 00-716 Warsaw, Poland. email:amss@camk.edu.pl\label{inst1} \and
	Astrophysics Group, Lennard-Jones Laboratories, Keele University, Keele, Staffordshire, ST5 5BG, UK	\label{inst2} \and
	Department of Physics, University of Warwick, Coventry CV4 7AL, UK\label{inst3}\and
	 Aix Marseille Universit\'e, CNRS, LAM (Laboratoire d'Astrophysique de Marseille) UMR 7326, 13388, Marseille, France\label{inst4}\and
	 INAF - Osservatorio Astrofisico di Torino, via Osservatorio 20, I-10025 Pino Torinese, Italy\label{inst5}\and
	SUPA, School of Physics and Astronomy, University of St. Andrews, North Haugh, St. Andrews, Fife, KY16 9SS, UK\label{inst6}\and
	Institut d'Astrophysique et de G\'{e}ophysique, Universit\'{e} de Li\`{e}ge, All\'{e}e du 6 Ao\^{u}t 17, Sart Tilman, Li\`{e}ge 1, Belgium\label{inst7}\and
	Institut d'Astrophysique de Paris, UMR7095 CNRS, Universit\'{e} Pierre \& Marie Curie, 98bis boulevard Arago, 75014, Paris, France\label{inst8}\and
	Observatoire de Haute-Provence, CNRS/OAMP, 04870 Saint-Michel-l'Observatoire, France\label{inst9}\and
	Observatoire de Gen\`{e}ve, Universit\'{e} de Gen\`{e}ve, 51 Chemin des Maillettes, 1290 Sauverny, Switzerland\label{inst10}\and
	Cavendish Laboratory, J J Thomson Avenue, Cambridge CB3 0HE, UK\label{inst11}\and
	Kavli Institute for Astrophysics \& Space Research, Massachusetts Institute of Technology, Cambridge, MA 02139, USA\label{inst12}
             }

   \date{}

\abstract{We report the discovery from the WASP survey of two exoplanetary systems, each consisting of a Jupiter-sized planet transiting an 11th magnitude ($V$) main-sequence star. WASP-104b orbits its star in 1.75~d, whereas WASP-106b has the fourth-longest orbital period of any planet discovered by means of transits observed from the ground, orbiting every 9.29~d. Each planet is more massive than Jupiter (WASP-104b has a mass of $1.27\pm0.05$~\mjup, while WASP-106b has a mass of $1.93\pm0.08$~\mjup). Both planets are just slightly larger than Jupiter, with radii of $1.14\pm0.04$ and $1.09\pm0.04$~\rjup for WASP-104 and WASP-106 respectively. No significant orbital eccentricity is detected in either system, and while this is not surprising in the case of the short-period WASP-104b, it is interesting in the case of WASP-106b, because many otherwise similar planets are known to have eccentric orbits.

}


   \maketitle

\section{Introduction}

Transiting planets are vital for our understanding of planetary systems, as planetary radii and absolute masses can be measured. Recently, the large number of planets and planetary candidates discovered by the {\it Kepler} satellite has extended the parameter space of transiting planet discovery and led to major advances in our understanding of the statistics of the Galactic planetary population. Planets transiting bright stars are required to increase our knowledge of the range of properties exhibited by the nearby planetary population, as well as to make advances in our understanding of planetary formation and evolution. Examples of planet characterisation that require bright target stars include high-precision radial velocity measurements to measure the orbital obliquity, and measurement of planetary transmission and emission spectra to infer atmospheric properties (e.g. \citealt{Winn_book}).

The Wide Angle Search for Planets (WASP; \citealt{Pollacco06}) and other wide-field ground-based surveys such as HAT-net \citep{bakos02} are the leading discoverers of bright ($8.5 \lesssim V \lesssim 12.5$) transiting systems, with around 150 planets between them.

Only a handful of the nearly one hundred planets discovered by WASP have orbital periods, $P$, greater than 5~d. This is because there is an intrinsic pile-up of planets at 3--5 days, and hence a relative dearth of planets at longer periods. This is further exacerbated by the fact that at long periods the probability that a planetary system is aligned such that transits are visible from Earth is significantly reduced. Furthermore, because a large number of transits must be observed to overcome the correlated noise present in wide-field survey photometry (e.g. \citealt{Smith1}), very long observational baselines are usually required to find longer-period systems.

Recently \cite{w117} reported the discovery of WASP-117b, the first WASP discovery with $P > 9$~d. Only two other systems with $P > 9$~d have been discovered from the ground by means of transits: HAT-P-15b \citep{hat15} and HAT-P-17b \citep{hat17}. Here we report the discovery of the fourth such planet, WASP-106b ($P = 9.29$~d), along with a much shorter-period planet, WASP-104b ($P = 1.75$~d). Both planets orbit $V=11$ stars lying in the constellation Leo.

\section{Observations}
\subsection{WASP photometry}

Both WASP-104 and WASP-106 were observed for four seasons by SuperWASP-N at the Observatorio del Roque de los Muchachos on La Palma, Spain, from 2008 February 05 to 2011 March 29. Each of the targets was also observed for two seasons by WASP-South at the South African Astronomical Observatory near Sutherland, South Africa, from 2009 January 16 to 2010 May 30. These observations resulted in 24632 data points on WASP-104 (13993 from SuperWASP-N and 10639 from WASP-South) and 23624 measurements of WASP-106 (13101 from SuperWASP-N and 10523 from WASP-South).

The WASP instruments consist of eight Canon 200mm f/1.8 lenses, each equipped with an Andor $2048\times2048$ e2v CCD camera, on a single robotic mount. Further details of the instrument, survey and data reduction procedures are described in \cite{Pollacco06} and details of the candidate selection procedure can be found in \cite{Cameron-etal07} and \cite{wasp3}.

The data revealed the presence of transit-like signals with a period of around 1.75 days in the case of WASP-104, and around 9.29 days for WASP-106. The combined WASP light curves are shown binned and folded on the best-fitting orbital periods in the upper panels of Figures \ref{fig:w104} and \ref{fig:w106}. The light curves were also searched for additional transiting components with periods up to 120~d, using the method described in \cite{Smith3}, but no such signals were found.

\subsection{Spectroscopic follow-up}

Spectroscopic observations of both targets were conducted using the SOPHIE spectrograph mounted on the 1.93-m telescope of the Observatoire de Haute-Provence, France, and with CORALIE on the 1.2-m Euler-Swiss telescope at La Silla, Chile. A total of 21 observations (10 with CORALIE, 11 with SOPHIE) were made of WASP-104 between 2013 January 08 and 2013 June 25. WASP-106 was observed 29 times (20 times with CORALIE, 9 with SOPHIE) between 2013 January 06 and 2014 February 08.

SOPHIE was used in high-efficiency ($R = 40000$) mode, with the CCD in slow-readout mode. A second fibre was placed on the sky to check for sky background contamination in the spectra of the target. Corrections of up to a few tens of m~s$^{-1}$ were applied to some exposures contaminated by moonlight. More information about the SOPHIE instrument can be found in \cite{sophie}. The CORALIE observations were conducted only in dark time to avoid moon light entering the fibre.

\begin{figure} 
\includegraphics[width=0.5\textwidth]{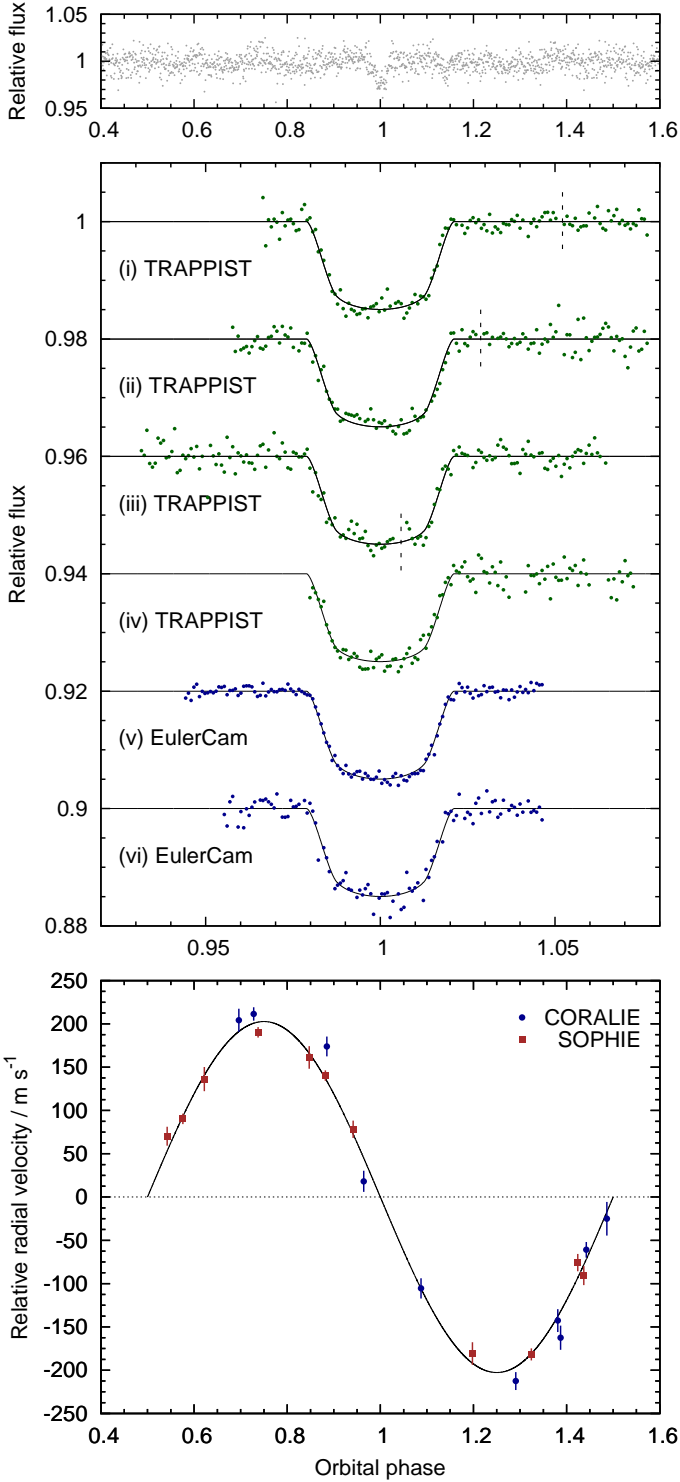} 
\caption{WASP-104 photometry and radial velocities. {\it Upper panel:} Phase-folded photometry from SuperWASP-N and WASP-South. Data are binned in phase with a bin-width equal to 120~s.
{\it Middle panel:} High-precision follow-up photometry. Each light curve is offset in flux for clarity, and binned using a 120~s bin width. The light curves are identified with the same numbers as those in Table \ref{tab:fup}. Our best-fitting model is plotted with a solid line and TRAPPIST's meridian flips are marked with dashed vertical lines.
{\it Lower panel:} Radial-velocity measurements from CORALIE and SOPHIE, along with our best-fitting MCMC solution (solid line). The jitter added to the radial velocity uncertainties is not shown. The centre-of-mass velocity, $\gamma$ = 28.548042 km s$^{-1}$, has been subtracted, as has the fitted offset $\gamma_{\rm COR-SOPH} = 280.23$~m~s$^{-1}$ between the two datasets. 
}
\label{fig:w104} 
\end{figure}

\begin{figure} 
\includegraphics[width=0.5\textwidth]{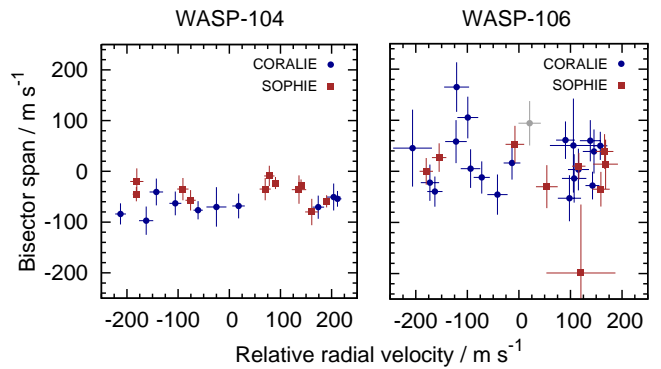} 
\caption{Radial velocity bisector span vs. relative radial velocity for WASP-104 (left) and WASP-106 (right). The indicated uncertainties in the radial velocities do not include the added jitter, and the uncertainties in the bisector span are taken to be twice the uncertainty in the radial velocity. The CORALIE point taken during transit and excluded from our MCMC analysis is shown in grey.
}
\label{fig:bisectors} 
\end{figure}

Observations of thorium-argon emission line lamps were used to calibrate both the SOPHIE and CORALIE spectra. The data were processed using the standard SOPHIE and CORALIE data-reduction pipelines. The resulting radial velocity data are listed in Table \ref{tab:rv} and plotted in the lower panels of Figures \ref{fig:w104} and \ref{fig:w106}. In order to rule out the possibility that either system is a blended eclipsing binary, or that the RV variation is a result of stellar activity, we examined the bisector spans (e.g. \citealt{Queloz01}). For both stars, the bisector spans exhibit no significant correlation with radial velocity (Figure \ref{fig:bisectors}), as expected for true planetary systems.

\begin{table}[h]
\centering
\caption{Radial velocities of WASP-104 and WASP-106}
\begin{tabular}{ccccrc} \hline
Star & BJD(UTC) & RV & $\sigma_{\mathrm RV}$ & BS~~ & Inst.$\dagger$\\
&-- 2450000& km s$^{-1}$ &  km s$^{-1}$ & km s$^{-1}$ & \\
 \hline
WASP-104 & 6364.6606  &  28.386  &  0.014  &  -0.082  & COR.\\
WASP-104 & 6365.6744  &  28.566  &  0.012  &  -0.053  & COR.\\
WASP-104 & 6368.7145  &  28.752  &  0.013  &  -0.030  & COR.\\
WASP-104 & 6371.6720  &  28.405  &  0.013  &  -0.027  & COR.\\
WASP-104 & 6410.4762  &  28.523  &  0.019  &  -0.057  & COR.\\
WASP-104 & 6438.4853  &  28.487  &  0.009  &  -0.062  & COR.\\
WASP-104 & 6442.4977  &  28.760  &  0.008  &  -0.040  & COR.\\
WASP-104 & 6444.5289  &  28.722  &  0.011  &  -0.055  & COR.\\
WASP-104 & 6450.5074  &  28.336  &  0.010  &  -0.069  & COR.\\
WASP-104 & 6469.4597  &  28.443  &  0.012  &  -0.048  & COR.\\
&&&&&\\
WASP-104 & 6300.5794  &  28.968  &  0.006  &  -0.028  & SOPH.\\
WASP-104 & 6329.6172  &  28.752  &  0.010  &  -0.057  & SOPH.\\
WASP-104 & 6330.5262  &  28.906  &  0.010  &  -0.009  & SOPH.\\
WASP-104 & 6331.5802  &  28.898  &  0.011  &  -0.035  & SOPH.\\
WASP-104 & 6361.4814  &  28.918  &  0.006  &  -0.023  & SOPH.\\
WASP-104 & 6363.5205  &  29.018  &  0.006  &  -0.059  & SOPH.\\
WASP-104 & 6377.3614  &  28.964  &  0.014  &  -0.036  & SOPH.\\
WASP-104 & 6378.3721  &  28.647  &  0.013  &  -0.020  & SOPH.\\
WASP-104 & 6401.4142  &  28.646  &  0.007  &  -0.045  & SOPH.\\
WASP-104 & 6403.3674  &  28.737  &  0.011  &  -0.035  & SOPH.\\
WASP-104 & 6423.3975  &  28.989  &  0.013  &  -0.080  & SOPH.\\
&&&&&\\
WASP-106 & 6364.6843  &  17.127  &  0.024  &  0.165  & COR.\\
WASP-106 & 6365.7257  &  17.206  &  0.020  &  -0.045  & COR.\\
WASP-106 & 6366.6264  &  17.353  &  0.046  &  0.051  & COR.\\
WASP-106 & 6368.7463  &  17.386  &  0.020  &  0.060  & COR.\\
WASP-106 & 6369.7726  &  17.363  &  0.020  &  0.004  & COR.\\
WASP-106 & 6370.7735  &  17.268  &  0.022  &  0.094  & COR.\\
WASP-106 & 6371.5952  &  17.154  &  0.019  &  0.005  & COR.\\
WASP-106 & 6371.7690  &  17.148  &  0.020  &  0.105  & COR.\\
WASP-106 & 6372.7689  &  17.075  &  0.018  &  -0.022  & COR.\\
WASP-106 & 6373.7840  &  17.125  &  0.021  &  0.058  & COR.\\
WASP-106 & 6410.5005  &  17.041  &  0.038  &  0.045  & COR.\\
WASP-106 & 6423.5873  &  17.390  &  0.014  &  -0.028  & COR.\\
WASP-106 & 6424.5533  &  17.405  &  0.014  &  0.050  & COR.\\
WASP-106 & 6438.5133  &  17.084  &  0.015  &  -0.040  & COR.\\
WASP-106 & 6441.4779  &  17.345  &  0.023  &  -0.053  & COR.\\
WASP-106 & 6443.5126  &  17.393  &  0.021  &  0.039  & COR.\\
WASP-106 & 6449.5284  &  17.234  &  0.016  &  0.017  & COR.\\
WASP-106 & 6469.4842  &  17.338  &  0.018  &  0.061  & COR.\\
WASP-106 & 6481.4663  &  17.355  &  0.024  &  -0.014  & COR.\\
WASP-106 & 6696.7609  &  17.175  &  0.016  &  -0.012  & COR.\\
&&&&&\\
WASP-106 & 6298.6199  &  17.033  &  0.014  &  0.027  & SOPH.\\
WASP-106 & 6329.6333  &  17.242  &  0.021  &  -0.030  & SOPH.\\
WASP-106 & 6330.5696  &  17.347  &  0.017  &  -0.035  & SOPH.\\
WASP-106 & 6331.5970  &  17.354  &  0.017  &  0.039  & SOPH.\\
WASP-106 & 6360.4915  &  17.304  &  0.013  &  0.010  & SOPH.\\
WASP-106 & 6363.5345  &  17.008  &  0.013  &  -0.000  & SOPH.\\
WASP-106 & 6377.3763  &  17.308  &  0.067  &  -0.199  & SOPH.\\
WASP-106 & 6378.3889  &  17.356  &  0.024  &  0.014  & SOPH.\\
WASP-106 & 6403.3926  &  17.180  &  0.018  &  0.053  & SOPH.\\

\hline
\\
\end{tabular}\\
$\dagger$ Spectrograph used. COR. = CORALIE on the Euler-Swiss telescope; SOPH. = SOPHIE on the 1.93-m telescope at the OHP.
\label{tab:rv}
\end{table}

\subsection{Photometric follow-up}

High-precision photometric follow-up observations were made from ESO's La Silla Observatory in Chile and from the Observatorio del Roque de los Muchachos on La Palma, Spain. The telescopes  used were the robotic Transiting Planets and Planetesimals Small Telescope (TRAPPIST; \citealt{TRAPPIST}) and the Euler-Swiss telescope with EulerCam \citep{EulerCam} at La Silla; and the Liverpool Telescope (LT) with the Rapid Imaging Search for Exoplanets camera (RISE; \citealt{rise2008, gibson2008}) and the Isaac Newton Telescope (INT) with the Wide Field Camera (WFC) on La Palma. A summary of these observations is given in Table \ref{tab:fup}. Due to the relatively long transit duration of over five hours, it proved difficult to observe a full transit of WASP-106b. This was, however, achieved using the LT on the night of 2014 February 25/26. Simultaneous observations from the nearby INT were stymied by dome-shutter problems during ingress, preventing observation of the full transit, although a partial transit was observed.

\begin{table} 
\caption{Observing log for follow-up transit photometry} 
\label{tab:fup} 
\begin{tabular}{@{\extracolsep{\fill}}clccc} 
\hline 
No.& Date/UT 			& Telescope		& Band 		&  full?$\dagger$ \\ 
\hline
\multicolumn{2}{l}{WASP-104:}&&&\\
(i) 	& 2013 Mar 29/30	& TRAPPIST	& $I + z$ 		& Y		\\
(ii)	& 2013 Apr 05/06	& TRAPPIST 	& $I + z$ 		& Y		\\
(iii)	& 2013 Apr 12/13	& TRAPPIST 	& $I + z$ 		& Y		\\
(iv)	& 2013 Jun 02/03	& TRAPPIST 	& $I + z$ 		& N		\\
(v)	& 2013 Apr 20		& Euler 		& $I$			& Y		\\
(vi)	& 2013 Apr 27		& Euler		& $I$			& Y		\\
\hline
\multicolumn{2}{l}{WASP-106:}&&&\\
(vii)	& 2013 May 13/14	& Euler 		& Gunn $r$	& N		\\
(viii)	& 2013 May 13/14	& TRAPPIST 	& $I + z$ 		& N		\\
(ix)	& 2014 Feb 25/26	& TRAPPIST 	& $I + z$ 		& N		\\
(x)	& 2014 Feb 25/26	& Liverpool	& $V + R$		& Y		\\
(xi)	& 2014 Feb 25/26	& Isaac Newton& $i'$		& N		\\
\hline 
\end{tabular}\\
$\dagger$Indicates whether or not the complete transit was observed
\end{table} 

\section{Determination of system parameters}
\subsection{Stellar parameters}
\label{sec:spec}

The individual CORALIE spectra were co-added to produce a single spectrum of each star with an average S/N of around 60:1 for WASP-104 and 100:1 for WASP-106. The analyses were performed using standard pipeline reduction products and the procedures given in \cite{Doyle13}. The projected stellar rotation velocities, \vsini, were determined by fitting the profiles of several unblended Fe~{\sc i} lines. We assumed macroturbulent velocities (\mactrb) of $2.5\pm0.6$~km~s$^{-1}$ for WASP-104 and $4.7\pm0.6$~km~s$^{-1}$ for WASP-106 using the asteroseismic-based calibration of \cite{Doyle14} and an instrumental resolution of 55000. The resultant stellar parameters are given in Table~\ref{tab:stellar}.

\subsubsection{Stellar activity}

We searched the WASP photometry for rotational modulations, using the sine-wave fitting algorithm described by \cite{w41}. We estimated the significance of periodicities by subtracting the fitted transit light curve and then repeatedly and randomly permuting the nights of observation. No significant periodic modulation was detected, to a 95 per cent confidence upper limit of 4 mmag in the case of WASP-104, and 1 mmag for WASP-106.

\subsubsection{Stellar age}
\label{sec:age}

\begin{figure} 
\includegraphics[width=0.5\textwidth]{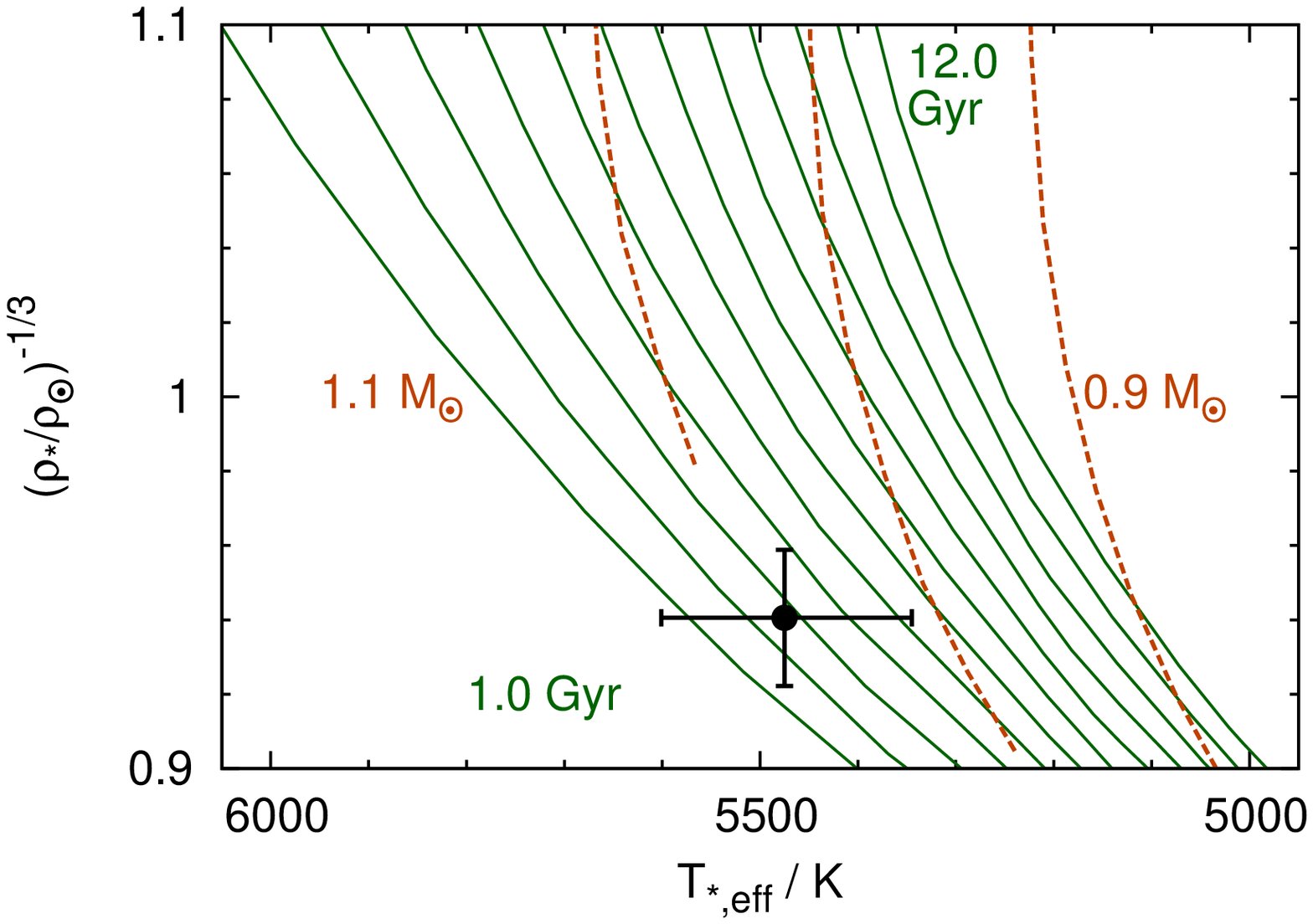} 
\caption{Modified Hertzsprung-Russell diagram for WASP-104 (black circle). Isochrones (solid green lines) from the models of \cite{Bressan12} are plotted for ages between 1.0 and 12.0 Gyr, spaced at 1.0 Gyr intervals. Also shown are evolutionary tracks (dashed orange lines) for 0.9, 1.0 and 1.1 \msol from \cite{Bertelli08}.}
\label{fig:iso104} 
\end{figure}

The low lithium abundance for WASP-104 suggests that the star is old (several Gyr, perhaps more than 5 Gyr; \citealt{2005A&A...442..615S}), and the low \vsini is consistent with this. From an analysis (Fig. \ref{fig:iso104}) performed using the isochrones of \cite{Bressan12}, we infer an age of around $3\pm2$~Gyr.

\begin{figure} 
\includegraphics[width=0.5\textwidth]{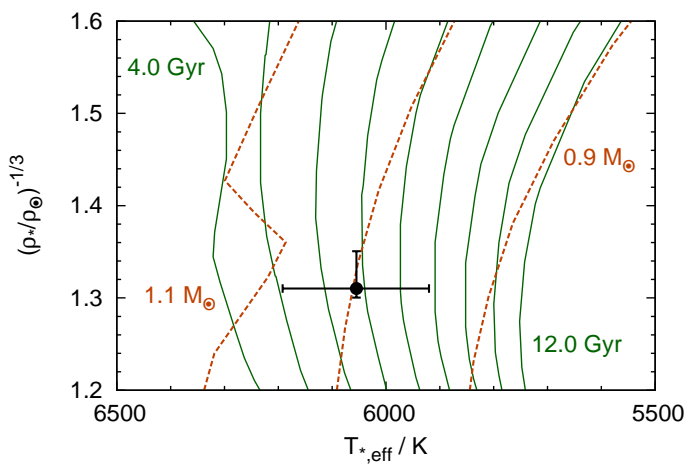} 
\caption{Modified Hertzsprung-Russell diagram for WASP-106 (black circle). Isochrones (solid green lines) from the models of \cite{Bressan12} are plotted for ages between 4.0 and 12.0 Gyr, spaced at 1.0 Gyr intervals. Also shown are evolutionary tracks (dashed orange lines) for 0.9, 1.0 and 1.1 \msol from \cite{Bertelli08}.}
\label{fig:iso106} 
\end{figure}

The lithium abundance of WASP-106 suggests an age of around 1 -- 2 Gyr. The \vsini gives an upper limit to the age of $1.4^{+0.7}_{-0.4}$~Gyr from the gyrochronological relation of \cite{2007ApJ...669.1167B}. Our isochrone analysis (Fig. \ref{fig:iso106}) gives an age estimate of $7\pm2$~Gyr.

Such discrepancies in the ages estimated from these two methods are not unusual, particularly for estimates derived from isochrone fitting (e.g. \citealt{Brown_ages}).

\subsubsection{Stellar distance}

We calculated the distances to WASP-104 and WASP-106 (see Table \ref{tab:stellar}) using Tycho apparent V-band magnitudes of $11.12\pm0.10$ and $11.21\pm0.11$ respectively. The stellar luminosities were calculated from our best-fitting stellar effective temperatures and radii (Table \ref{tab:mcmc}), and we adopted bolometric corrections (from \citealt{Flower96}) of $-0.15$ and $-0.05$, for WASP-104 and WASP-106 respectively. We use the interstellar Na D lines and the calibration of \cite{MZ97} to adopt colour excesses, $E(B-V)=0.04\pm0.01$ and $E(B-V)=0.02\pm0.01$ for WASP-104 and WASP-106 respectively, to correct for interstellar reddening.

\begin{table}[h]
\caption{Stellar parameters}
\begin{tabular}{lll} \hline
Parameter  & 		WASP-104 & 			WASP-106	\\	
\hline					
RA (J2000.0) & 	10h42m24.61s & 		11h05m43.13s		\\
Dec (J2000.0) & 	$+07^{\circ}~26\arcmin~06.3\arcsec $&$-05^{\circ}~04\arcmin~45.9\arcsec $		\\
Identifier &		USNO-B1.0 0974-0234922	& TYC 4927-1063-1 \\
\teff  / K    &  		5450 $\pm$ 130 & 	6000$\pm$150				\\
\logg (cgs)     & 		4.5 $\pm$  0.2 & 		4.0 $\pm$  0.2		\\	
\mictrb /  \kms  &  0.9  $\pm$ 0.1&  1.2  $\pm$ 0.1 \\
\mactrb / \kms & $2.5\pm0.6$ & $4.7\pm0.6$\\
\vsini    /  \kms   &  	0.4 $\pm$ 0.7 &		6.3 $\pm$ 0.7 			\\
$\log A$(Li)  &    	$< 0.7$ & 				2.45 $\pm$ 0.12	\\
Sp. Type   &   		G8 & 				F9			\\
Distance /pc  &    	143$\pm$10 & 			283$\pm$21		\\	
Mass  /$\msol$     &   1.02$\pm$0.09 &		1.27$\pm$0.15	\\	 
Radius  /$\rsol$   &   0.93$\pm$0.23 &		1.85$\pm$0.52	\\	
{[Fe/H]}   & 		$+$0.32 $\pm$ 0.09 &	$-$0.09 $\pm$ 0.09\\
$V$-mag (Tycho) & 	$11.12\pm0.10$ & 		$11.21\pm0.11$\\
\hline
\\
\end{tabular}
\label{tab:stellar}
\newline {\bf Note:} The spectral type was estimated from \teff\
using the table of \cite[p. 507]{gray-book}. The mass and radius were estimated using the \cite{2010A&ARv..18...67T} calibration.

\end{table}

\subsection{Planetary system parameters}

For each system the photometric data were combined with the radial velocities and analysed simultaneously using the Markov Chain Monte Carlo (MCMC) method. 

\begin{figure} 
\includegraphics[width=0.5\textwidth]{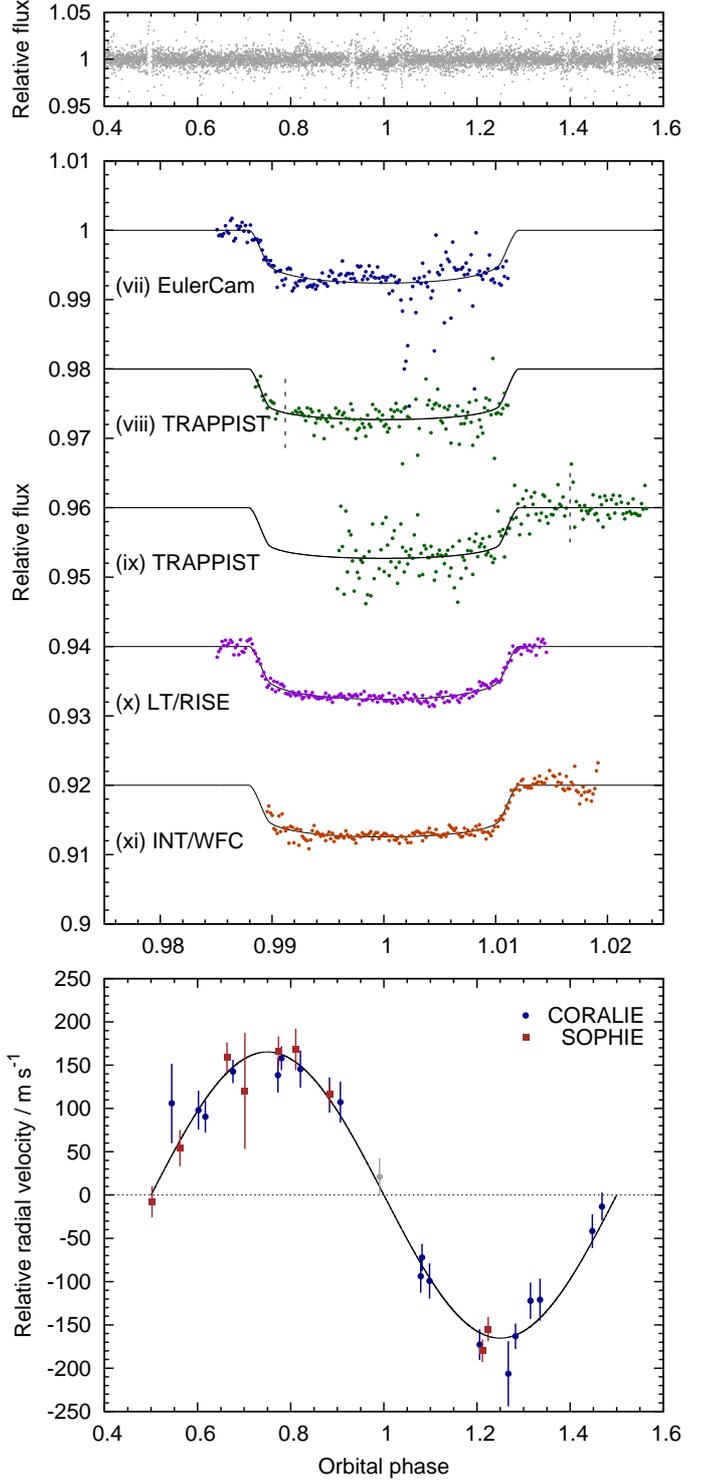} 
\caption{WASP-106 photometry and radial velocities. {\it Upper panel:} Phase-folded photometry from SuperWASP-N and WASP-South. Data are binned in phase with a bin-width equal to 120~s.
{\it Middle panel:} High-precision follow-up photometry. Each light curve is offset in flux for clarity, and binned using a 120~s bin width. The light curves are identified with the same numbers as those in Table \ref{tab:fup}. Our best-fitting model is plotted with a solid line and TRAPPIST's meridian flips are marked with dashed vertical lines.
{\it Lower panel:} Radial-velocity measurements from CORALIE and SOPHIE, along with our best-fitting MCMC solution (solid line). The CORALIE point taken during transit and excluded from our MCMC analysis is shown in grey. The centre-of-mass velocity, $\gamma$ = 17.24744 km s$^{-1}$, has been subtracted, as has the fitted offset $\gamma_{\rm COR-SOPH} = -59.57$~m~s$^{-1}$ between the two datasets. 
}
\label{fig:w106} 
\end{figure} 

We use the current version of the MCMC code described in \cite{Cameron-etal07} and \cite{wasp3}, which uses the relation between stellar mass, effective temperature, density and metallicity appropriate for stars with masses between 0.2 and 3.0 \msol from \cite{SWorth_homo4}. Briefly, the radial velocity data are modelled with a Keplerian orbit and the photometric transits are fitted using the model of \cite{M&A} and limb-darkening was accounted for using a four-coefficient, non-linear model, employing coefficients appropriate to the passbands from the tabulations of \cite{Claret, Claret04}. The coefficients were determined using an initial interpolation in $\log g_{*}$ and [Fe/H] (values from Table \ref{tab:stellar}), and an interpolation in $T_{\rm *, eff}$  at each MCMC step. The coefficient values corresponding to the best-fitting value of $T_{\rm *, eff}$ are given in Table \ref{tab:limb}.

The MCMC proposal parameters we use are: the epoch of mid-transit, $T_{\rm c}$; the orbital period, $P$; the transit duration, $T_{\rm 14}$; the fractional flux deficit that would be observed during transit in the absence of stellar limb-darkening, $\Delta F$; the transit impact parameter for a circular orbit, $b = a\cos(i_P)/\rstar$; the stellar reflex velocity semi-amplitude, $K_{\mathrm{*}}$; the stellar effective temperature, \teff, and the stellar metallicity, [Fe/H]. When fitting for a non-zero orbital eccentricity, there are an additional two proposal parameters: $\sqrt{e}\cos\omega$ and $\sqrt{e}\sin\omega$, where $e$ is the orbital eccentricity, and $\omega$ is the argument of periastron \citep{wasp30}. The best-fitting system parameters are taken to be the median values of the posterior probability distribution, and are recorded in Table \ref{tab:mcmc}.

\subsubsection{WASP-104}
\label{sec:w104fit}

There is a star of approximately equal brightness to WASP-104 (USNO-B1.0 R = 10.82, c.f. 11.28 for WASP-104), located 29\arcsec to the North. We found from our observations with the Euler Telescope that this neighbouring object (TYC 260-1073-1) is itself a visual binary, with the components separated from each other by around 1\arcsec. Further, we found this binary to be unrelated to WASP-104, which has a systemic radial velocity of 28 \kms compared to -3 \kms for the binary. This conclusion is supported by proper motions from the UCAC4 catalogue \citep{UCAC4}, which indicate that it is unlikely that WASP-104 ($\mu_\alpha = -30.3\pm2.7 \mathrm{~mas~yr^{-1}}$; $\mu_\delta = -2.3\pm1.6\mathrm{~mas~yr^{-1}}$) and the neighbouring object ($\mu_\alpha = -14.4\pm6.6 \mathrm{~mas~yr^{-1}}$; $\mu_\delta = 4.8\pm2.2\mathrm{~mas~yr^{-1}}$) are co-moving.

Although this object is easily excluded from the photometric apertures used in our follow-up observations, it falls within the 3.5-pixel (= 48\arcsec) WASP aperture, thus rendering the WASP photometry unreliable for measuring the transit depth. After determining the best-fitting orbital period from an MCMC analysis of all the available data, we exclude the WASP photometry from subsequent analysis. Instead, we fix the orbital period to the value previously determined and use MCMC to analyse only the follow-up photometry and the radial velocities.

An initial fit for an eccentric orbit gave $e =  0.0092^{+0.0123}_{-0.0066}$ (with $\omega = 15^{+106}_{-125}$~degrees). This is consistent with a circular orbit; according to the F-test of \cite{lucy_sweeney} there is a $>99$ per cent chance that the apparent eccentricity could have arisen if the underlying orbit were circular. We also tried fitting for a drift in the radial velocities by fitting for a linear function in time. Such a trend may be indicative of a third body in the system, but fitting for radial acceleration, $d\gamma/dt$, is clearly not justified in this case, because we obtain $d\gamma/dt = -24\pm28$~m~s$^{-1}$~yr$^{-1}$ (Figure \ref{fig:rv_drift}).  

Linear functions of time were fitted to each light curve at each step of the MCMC, to remove systematic trends. We added `jitter' terms of 15~m~s$^{-1}$ for CORALIE and 6~m~s$^{-1}$ for SOPHIE to the formal RV uncertainties in quadrature so as to obtain a spectroscopic reduced $\chi^2$ of unity.

\subsubsection{WASP-106}
\label{sec:w106fit}

One of the CORALIE radial velocity measurements was made during transit, and we exclude this point from our MCMC analysis because we do not model the Rossiter-McLaughlin effect. The excluded point is, however, plotted (in grey) in the lower panel of Figure \ref{fig:w106}. It was not necessary to add any `jitter' to the RV uncertainties.

\begin{figure} 
\includegraphics[width=0.5\textwidth]{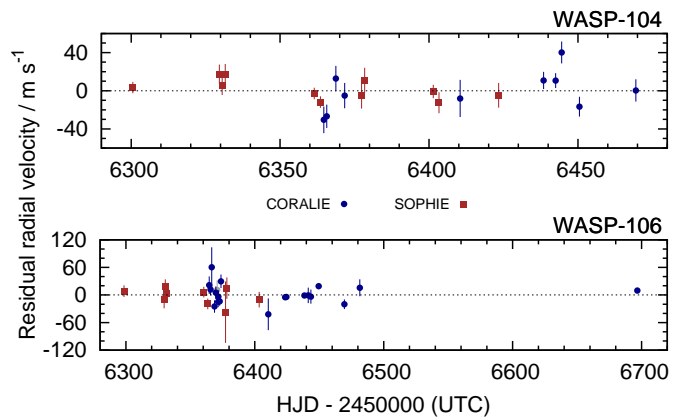} 
\caption{Residual radial velocity as a function of time for WASP-104 (upper) and WASP-106 (lower). There is no evidence of radial acceleration in either system. The indicated uncertainties in the radial velocities do not include the added jitter. The CORALIE point taken during transit and excluded from our MCMC analysis is shown in grey.
}
\label{fig:rv_drift} 
\end{figure}

Fitting for an eccentric orbit gives $e = 0.031^{+0.032}_{-0.022}$ (with $\omega = 103^{+30}_{-82}$~degrees). This is again consistent with a circular orbit (the \cite{lucy_sweeney} F-test probability is 12 per cent). The best-fitting radial acceleration is $d\gamma/dt = 11^{+13}_{-18}$~m~s$^{-1}$~yr$^{-1}$, which is also consistent with zero (Figure \ref{fig:rv_drift}). We therefore adopt, as for the WASP-104 system, a circular orbit and a constant systemic radial velocity.

Linear functions of time were fitted to each light curve at each step of the MCMC, to remove systematic trends. In the case of the LT light curve (x), the function fitted was quartic, rather than linear in time. A function of CCD position, airmass and sky background instead of time was fitted to the INT light curve (xi). In each case, the addition of extra terms was justified by an improved value of the Bayesian Information Criterion (BIC; \citealt{BIC}).

\begin{table*} 
\caption{System parameters for WASP-104 and WASP-106} 
\label{tab:mcmc}
\begin{tabular}{lcccc}
\hline
\hline
Parameter & Symbol & Unit & WASP-104 & WASP-106\\
\hline 
\textit{Model parameters:} &\\
&\\
Orbital period	    	    	    	    & 	$P$ & d & 	1.7554137$^{+ 0.0000018}_{- 0.0000036}\dagger$& $9.289715\pm0.000010$\\
Epoch of mid-transit	    	    	    & 	$T_{\rm c}$ &HJD, UTC & 	 2456406.11126$\pm 0.00012$& 2456649.54972$\pm0.00033$\\
Transit duration    	    	    	    & 	$T_{\rm 14}$ &d & 	 $0.07342\pm0.00056$& 0.22346$^{+ 0.00104}_{- 0.00090}$\\
Planet-to-star area ratio   	    	    & 	$\Delta F=R_{\rm P}^{2}$/R$_{*}^{2}$&... & 	 $0.01474\pm0.00020$& $0.00642\pm0.00018$\\
Transit impact parameter    	    & 	$b$ &...& $0.724\pm0.014$ &	0.127$^{+ 0.149}_{- 0.087}$\\
Stellar orbital velocity semi-amplitude     & 	$K_{\rm *}$ &m s$^{-1}$ & 	 $202.7\pm4.3$& $165.3\pm4.3$\\
System velocity     	    	    	    &     	$\gamma$ &km s$^{-1}$ & 	 $28.54804\pm0.00010$& $17.24744\pm0.00019$\\
Velocity offset between CORALIE and SOPHIE & $\gamma_{\rm COR-SOPH}$ & m s$^{-1}$ & 	 $280.2\pm1.0$& $-59.57\pm0.13$\\
Stellar effective temperature         &  $T_{\rm *, eff}$ & K & 	 $5475\pm127$& $6055\pm136$\\
Stellar metallicity			    &  [Fe/H] & dex & 	 $+0.320\pm0.090$& $-0.090\pm 0.090$\\
&\\	
\textit{Derived parameters:}\\	
&\\	
Ingress / egress duration    	    & 	$T_{\rm 12}=T_{\rm 34}$ &d & 	 $0.01535\pm0.00075$& 0.01685$^{+ 0.00118}_{- 0.00039}$\\
Orbital inclination angle   	    	    & 	$i_P$ &$^\circ$  & 	 $83.63\pm0.25$& 89.49$^{+ 0.35}_{- 0.64}$\\
Orbital eccentricity (adopted)	    	    	    & 	$e$ &...& 0&0 \\	 
Orbital eccentricity (3-$\sigma$ upper-limit) &... &...& 0.060& 0.13\\
Stellar mass	    	    	    	    & 	$M_{\rm *}$ & $M_{\rm \odot}$ & 	 $1.076\pm0.049$& $1.192\pm0.054$\\
Stellar radius	    	    	    	    & 	$R_{\rm *}$ & $R_{\rm \odot}$ & 	 $0.963\pm0.027$& 1.393$^{+ 0.048}_{- 0.028}$\\
log (stellar surface gravity)     	    & 	$\log g_{*}$ & (cgs) & 	 $4.503\pm0.016$& 4.226$^{+ 0.010}_{- 0.022}$\\
Stellar density     	    	    	    & 	$\rho_{\rm *}$ &$\rho_{\rm \odot}$ & 	 $1.207\pm0.070$& 0.445$^{+ 0.010}_{- 0.039}$\\
Planet mass 	    	    	    	    & 	$M_{\rm P}$ &$M_{\rm Jup}$ & 	 $1.272\pm0.047$& $1.925\pm0.076$\\
Planet radius	    	    	    	    & 	$R_{\rm P}$ &$R_{\rm Jup}$ & 	 $1.137\pm0.037$& 1.085$^{+ 0.046}_{- 0.028}$\\
log (planet surface gravity)     	    & 	$\log g_{\rm P}$ & (cgs) & 	 $3.353\pm0.023$& 3.571$^{+ 0.021}_{- 0.031}$\\
Planet density	    	    	    	    & 	$\rho_{\rm P}$ &$\rho_{\rm J}$ & 	 $0.866\pm0.071$& 1.50$^{+ 0.11}_{- 0.16}$\\
Scaled orbital major semi-axis     &   $a/R_{\rm *}$ &...& 	 $6.52\pm0.13$& 14.20$^{+ 0.11}_{- 0.43}$\\
Orbital major semi-axis     	    	    & 	$a$ &au  & 	 $0.02918\pm 0.00045$& $0.0917\pm 0.0014$\\
Planet equilibrium temperature    & 	$T_{\rm P, A=0}$ &K & 	 $1516\pm39$& $1140\pm29$\\
~~~~~(uniform heat redistribution)&&&&\\
\hline 
\end{tabular} \\ 
$\dagger$The orbital period of WASP-104b and its uncertainty was determined from a fit to all the data (including photometry from the WASP instruments). The period was fixed in subsequent analyses (which excluded the WASP photometry). See Sec \ref{sec:w104fit}.\\
The following constant values are used: au $= 1.49598\times10^{11}$~m; $R_{\rm \odot} = 6.9599\times10^8$~m; $M_{\rm \odot} = 1.9892\times10^{30}$~kg;\\
$R_{\rm Jup} = 7.1492\times10^7$~m; $M_{\rm Jup} = 1.89896\times10^{27}$~kg; $\rho_{\rm J} = 1240.67$~kg m$^{-3}$. \\ 
\end{table*} 

\begin{table*}
\centering
\caption{Stellar limb-darkening coefficients}
\label{tab:limb}
\begin{tabular}{llllrrrr} \hline 	
Star 		& Instruments 		& Observation bands 			& Claret band & $a_1$ &  $a_2$  &  $a_3$ & $a_4$  \\
\hline 
WASP-104	& TRAPPIST		& Cousins $I$ + Sloan $z^\prime$	& Sloan $z^\prime$ 		& 0.821 & $-0.828$ & 1.212 & $-0.527$\\
WASP-104	& Euler			& Cousins $I$					& Cousins $I$			& 0.821 & $-0.835$ & 1.308 & $-0.573$\\
WASP-106 	& WASP / Euler / LT	& Broad ( 400 -- 700 nm) / Gunn $r$	/ $V + R$ & Cousins $R$ 	&0.497 & 0.273  & 0.034 & $-0.096$\\
WASP-106 	& TRAPPIST		& Cousins $I$ + Sloan $z^\prime$	& Sloan $z^\prime$ 		& 0.592 & $-0.134$ & 0.337 & $-0.199$\\
WASP-106	& INT			& Sloan $i^\prime$				& Sloan $i^\prime$		& 0.559 & 0.033  & 0.223 & $-0.162$\\
\hline 
\end{tabular}\\
\end{table*} 

\section{Results and Discussion}

We find that WASP-104 and WASP-106 are both main-sequence stars. WASP-104 has a mass and radius similar to that of the Sun, while WASP-106 is around 20 per cent more massive than the Sun, and has a radius around 40 per cent larger. Both planets are more massive than Jupiter (WASP-104b has a mass of $1.27\pm0.05$~\mjup, while WASP-106b has a mass of $1.93\pm0.08$~\mjup). WASP-104b and WASP-106b are not bloated like many hot Jupiters. Indeed with radii of $1.14\pm0.04$ and $1.09\pm0.04$~\rjup respectively, neither planet is much larger than Jupiter.

\subsection{The circular nature of the orbit of WASP-106b}

It is interesting to compare WASP-106 to WASP-117 \citep{w117}, a recently discovered transiting system consisting of a hot Jupiter in a 10.0~d orbit that is both significantly eccentric ($e = 0.302 \pm0.023$) and mis-aligned ($\psi = 69.6^{+4.7}_{-4.1}$~deg; $\beta = -44\pm11$~deg). By contrast, WASP-106b orbits in a circular or near-circular orbit (Sec. \ref{sec:w106fit}). The stars WASP-106 and WASP-117 ($M_{\rm *} = 1.126\pm0.029 M_{\rm \odot}$, $R_{\rm *} = 1.170^{+ 0.067}_{- 0.059} R_{\rm \odot}$, $T_{\rm *, eff} = 6038\pm88$~K) are also very similar to each other.

Such planets may offer important evidence as to which of two proposed planetary migration scenarios is dominant. Disk-driven migration should produce giant planets in circular orbits with low obliquities, whereas  planet-planet scattering and migration through the Kozai mechanism should lead to highly eccentric orbits, with large obliquities (see \citealt{w117} and references therein). These eccentric orbits may be subsequently circularised through tidal interactions with the star.

The vast majority of known hot Jupiters lie in circular or near-circular orbits, and distinguishing between established circular orbits and once-eccentric orbits that have been relatively quickly made circular by the huge tidal forces at play is non-trivial. Hot Jupiters with periods longer than 5~d, where the observed pile-up lies, may offer a clue to this puzzle, because the tidal forces acting to circularise the orbit are lesser.

Using Equation (1) of \cite{Jackson08}, we calculate the circularisation time-scale, $\tau_e = \left(\frac{1}{e} \frac{de}{dt}\right)^{-1}$, for the orbit of WASP-106b, in terms of the tidal dissipation parameters for the planet, $Q_{\rm P}$, and for the star, $Q_{\rm *}$,
\begin{equation}
\tau_e = \left(\frac{0.046}{\left (\frac{Q_{\rm P}}{10^{5.5}}\right)} + \frac{0.0023}{\left (\frac{Q_{\rm *}}{10^{6.5}}\right)}\right)^{-1}~\mathrm{Gyr,}
\end{equation}
Adopting $Q_{\rm P} = 10^{5.5}$ and $Q_{\rm *} = 10^{6.5}$ (the best-fitting values from the study of \citealt{Jackson08}), we find $\tau_e = 20.7$ Gyr \footnote{For comparison, the equivalent value for WASP-104 is just 7.6 Myr.}. Even in the extreme case that $Q_{\rm P} = Q_{\rm *} = 10^{5}$, the circularisation time-scale is still as long as 4.6~Gyr, comparable to the age of the system. Hence it seems probable that the orbit of WASP-106b has not been circularised from a highly eccentric starting point, but rather that the orbit has remained close to circular for the lifetime of the system.

A large fraction of similar planets, however, lie on significantly eccentric orbits. Using the {\it Exoplanets Data Explorer} \citep{exoplanets.org}, we find that there are 34 known planets of at least half the mass of Jupiter with orbital periods in the range $9 \le P \le 100$~d (including WASP-106b). Around half of these (18) are known to have significantly eccentric ($e \ge 0.1$) orbits, while the remainder have near-circular orbits, or their eccentricity is unknown.

It would be interesting to see if the obliquity of the WASP-106 system is also low, as well as the eccentricity. Measuring the sky-projected angle between the stellar spin axis and the orbital axis through the Rossiter-McLaughlin effect would give an insight into this. Such a measurement should prove possible; we predict an amplitude of around 26~m~s$^{-1}$ for the Rossiter-McLaughlin effect, and a typical RV precision of 4~m~s$^{-1}$ per 10-minute exposure with HARPS.

\begin{acknowledgements}

We acknowledge the significant contributions made to the construction and operation of WASP-South by Coel Hellier, who also identified WASP-104 and WASP-106 as candidate planetary systems.

WASP-South is hosted by the SAAO and SuperWASP by the Isaac Newton Group and the Instituto de Astrof\'{i}sica de Canarias;  we gratefully acknowledge their ongoing support and assistance. Funding for WASP comes from consortium universities and from the UK's Science and Technology Facilities Council (STFC).  TRAPPIST is a project funded by the Belgian Fund for Scientific Research (FNRS) with the participation of the Swiss National Science Foundation. MG and EJ are FNRS Research Associates. The Liverpool Telescope is operated on the island of La Palma by Liverpool John Moores University in the Spanish Observatorio del Roque de los Muchachos of the Instituto de Astrof\'{i}sica de Canarias with financial support from the UK STFC. The Isaac Newton Telescope is operated on the island of La Palma by the Isaac Newton Group in the Spanish Observatorio del Roque de los Muchachos of the Instituto de Astrof\'{i}sica de Canarias; observations were made under programme I/2014A/09. AHMJT is a Swiss National Science Foundation fellow under grant number P300P2-147773.
\end{acknowledgements}

\bibliographystyle{aa}
\bibliography{iau_journals,refs2}

\end{document}